\documentclass[runningheads]{llncs}

\usepackage[export]{adjustbox}
\usepackage{hyperref}
\usepackage{booktabs}
\usepackage{graphicx}
\usepackage{subcaption}
\usepackage{cancel}
\usepackage{amsmath,amsfonts,amssymb,mathtools} 
\usepackage[misc,geometry]{ifsym} 
\usepackage[utf8]{inputenc}

\begin{document}

\title{Voxel-level Importance Maps for Interpretable Brain Age Estimation}

\author{Kyriaki-Margarita Bintsi \inst{1} \textsuperscript{(\Letter)}, Vasileios Baltatzis \inst{1, 2}, Alexander Hammers \inst{2}, Daniel Rueckert \inst{1,3}}
\authorrunning{K. M. Bintsi et al.}

\institute{BioMedIA, Department of Computing, Imperial College London, UK \\
\email{m.bintsi19@imperial.ac.uk}\\ 
\and Biomedical Engineering and Imaging Sciences, King's College London, UK \\
\and Technical University of Munich, Germany
}

\maketitle              
\begin{abstract}
Brain aging, and more specifically the difference between the chronological and the biological age of a person, may be a promising biomarker for identifying  neurodegenerative diseases. For this purpose accurate prediction is important but the localisation of the areas that play a significant role in the prediction is also crucial, in order to gain clinicians' trust and reassurance about the performance of a prediction model.
Most interpretability methods are focused on classification tasks and cannot be directly transferred to regression tasks. In this study, we focus on the task of brain age regression from 3D brain Magnetic Resonance (MR) images using a Convolutional Neural Network, termed prediction model. We interpret its predictions by extracting importance maps, which discover the parts of the brain that are the most important for brain age. In order to do so, we assume that voxels that are not useful for the regression are resilient to noise addition. We implement a noise model which aims to add as much noise as possible to the input without harming the performance of the prediction model. We average the importance maps of the subjects and end up with a population-based importance map, which displays the regions of the brain that are influential for the task. We test our method on 13,750 3D brain MR images from the UK Biobank, and our findings are consistent with the existing neuropathology literature, highlighting that the hippocampus and the ventricles are the most relevant regions for brain aging.

\end{abstract}

\section{Introduction}

Alzheimer's disease (AD) is the most common cause of dementia \cite{AlzheimersAssociation2019}. AD leads to the atrophy of the brain more quickly than healthy aging and is a progressive neurodegenerative disease, meaning that more and more parts of the brain are damaged over time. The atrophy primarily appears in brain regions such as hippocampus, and it afterwards progresses to the cerebral neocortex. At the same time, the ventricles of the brain as well as cisterns, which are outside of the brain, enlarge \cite{Savva2009}.

Healthy aging also results in  changing of the brain, following specific patterns \cite{Alam2014}. Therefore, a possible biomarker used in AD is the estimation of the brain (biological) age of a subject which can then be compared with the subject's real (chronological) age \cite{Franke2019}. People at increased risk can be identified by the deviation between these two ages and early computer-aided detection of possible patients with neurodegenerative disorders can be accomplished. For this reason, a large body of research has focused on estimating brain age, especially using Magnetic Resonance (MR) images, which have long been used successfully in the measurement of brain changes related to age \cite{Good2001}. 


Recently, deep learning models have proved to be successful on the task of brain age estimation, providing relatively high accuracy. They are designed to find correlation and patterns in the input data and, in the supervised learning setting, associate that with a label, which in our case is the age of the subject. The models are trained on a dataset of MR images of healthy brains, estimating the expected chronological age of the subject. During training, the difference of the chronological age and the predicted age needs to be as small as possible. During test time, an estimated brain age larger than the subject's chronological age indicates accelerated aging, thus pointing to a possible AD patient.

\begin{figure}[t]
\centering
\includegraphics[width = \textwidth]{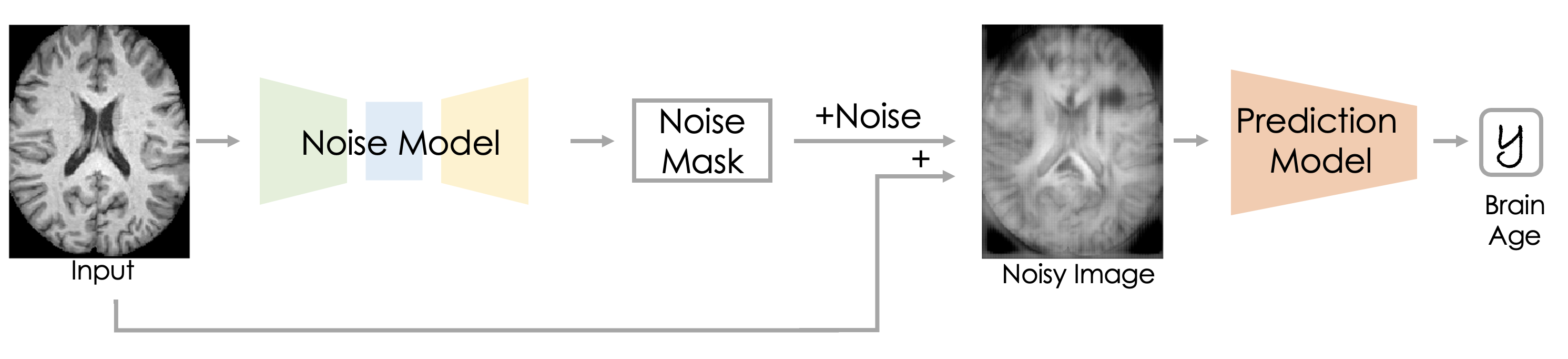}
\caption{Overview of the proposed idea. A noise model is trained with the purpose of adding as much noise as possible to the input. The output is a noise mask, in which noise sampled from the standard normal distribution is added. The result is then added to the input image and is used as an input to a pretrained prediction model with frozen parameters. The aim is to create a noisy image that will maximize the noise level while also not harming the performance of the prediction model.} \label{fig1}
\end{figure}

Convolutional Neural Networks (CNNs) are used with the purpose of an accurate brain age estimation while using the minimum amount of domain information since they can process raw image data with little to no preprocessing required. Many studies provide very accurate results, with mean absolute error (MAE) as low as around 2.2 years \cite{Peng2021,Pardakhti2020,Liu2020}. However, most of these approaches purely focus on minimization of the prediction error while considering the network as a black box. Recent studies have started to try to identify which regions are most informative for age prediction. For example, in \cite{Bintsi2020} the authors provided an age prediction for every patch of the brain instead of whole brain. Although the predictions and results presented in \cite{Bintsi2020} were promising, and the localisation was meaningful, the use of large patches meant that the localisation was not very precise. Similar approaches has been explored, such as \cite{Ballester2021} in which slices of the brain were used instead of patches. In \cite{Popescu}, the authors provided an age estimation per voxel but the accuracy of the voxel-wise estimations dropped significantly. In \cite{Levakov2020},  an ensemble of networks and gradient-based techniques \cite{Smilkov2017} were used in order to provide population-based explanation maps. Finally, 3D explanation maps were provided in \cite{Bass2021}, but the authors used image translation techniques and generative models, such as VAE-GAN networks.

In computer vision, a common approach to investigate black-box models such as CNNs is to use saliency maps, which show which regions the network focuses on, in order to make the prediction. An overview of the existing techniques for explainability of deep learning models in medical image analysis may be found in \cite{Singh}. Gradient-based techniques, such as guided backpropagation \cite{Springenberg2015}, and Grad-CAM \cite{Selvaraju2020}, make their conclusions based on the gradients of the networks with respect to a given input. Grad-CAM is one of the most extensively used approach and usually results in meaningful but very coarse saliency maps. Gradient-based techniques are focused on classification and to our knowledge they do not work as expected for regression task because they detect the features that contributed mostly to the prediction of a specific class. Occlusion-based techniques \cite{Zeiler2014} have also been widely explored in the literature and they can be used both for classification and regression tasks. The idea behind occlusion techniques is very simple: The performance of the network is explored after hiding different patches of the images, with the purpose of finding the ones that affect the performance the most. It is a promising and straightforward approach but bigger patches provide coarse results. On the other hand, the smaller the patches, the greater the computational and time constraints are, which can be a burden in their application. 

A recent approach which leverages the advantages of occlusion techniques while also keeping computational and time costs relatively low is U-noise \cite{Koker} which uses the addition of noise in the images pixel-wise, while keeping the performance of the network unchanged with the purpose of understanding where the deep learning models focus in order to do their predictions. The authors created interpretability maps for the task of pancreas segmentation using as input 2D images using noise image occlusion. In this paper, inspired by the work described above \cite{Koker} and the idea that when a voxel is not important for the task, then the addition of noise on this specific voxel will not affect the performance of our network, we make the following contributions: 
1) We adapt the architecture of U-noise (Figure \ref{fig1}), which was originally used for pancreas segmentation, for the task of brain age regression and visualise the parts of the brain that played the most important role for the prediction by training a noise model that dynamically adds noise to the image voxel-wise, providing localised and fine-grained importance maps.
2) We extend the U-noise architecture to 3D instead of 2D to accommodate training with three-dimensional volumetric MR images;
3) We propose the use of an autoencoder-based pretraining step on a reconstruction task to facilitate faster convergence of the noise model;
4) We provide a population-based importance map which is generated by aggregating subject-specific importance maps and highlights the regions of the brain that are important for healthy brain aging.

\section{Materials \& Methods}

\subsection{Dataset \& Preprocessing}
We use the UK Biobank (UKBB) dataset \cite{Alfaro-Almagro2018} for estimating brain age and extracting the importance maps. UKBB contains a broad collection of brain MR images, such as T1-weighted, and T2-weighted. The dataset we use in this work includes T1-weighted brain MR images from around 15,000 subjects. The images are provided by UKBB skull-striped and non-linearly registered to the MNI152 space. After removing the subjects lacking the information of age, we end up with 13,750 subjects with ages ranging from 44 to 73 years old, 52,3\% of whom are females and 47.7\% are males. The brain MR images have a size of 182x218x182 but we resize the 3D images to 140x176x140 to remove a large part of the background and at the same time address the memory limitations that arise from the use of 3D data, and normalise them to zero mean and unit variance.

\subsection{Brain Age Estimation}
We firstly train a CNN, the prediction model, $f_{\theta}$, with parameters $\theta$, for the task of brain age regression. We use the 3D brain MR images as input to 3D ResNet-18, similar to the one used in \cite{Bintsi2020}, which uses 3D convolutional layers instead of 2D \cite{He2016}. The network is trained with a Mean Squared Error (MSE) loss and its output is a scalar representing the predicted age of the subject in years. 

\subsection{Localisation}

An overview of the proposed idea is shown in Figure \ref{fig1}.

\subsubsection{U-Net Pretraining}
The prediction model and the noise model have identical architectures (2D U-Net \cite{Ronneberger2015U-net:Segmentation}) in \cite{Koker}, as they are both used for image-to-image tasks, which are segmentation and noise mask generation, respectively. Therefore, in that case, the noise model is initialized with the weights of the pretrained prediction model. However, in our case, the main prediction task is not an image-to-image task but rather a regression task and thus, the prediction model's (a 3D ResNet as described above) weights cannot be used to initialize the noise model. Instead, we propose to initially use the noise model $f_{\psi}$ with parameters $\psi$, which has the architecture of a 3D U-Net as a reconstruction model, for the task of brain image reconstruction. By pretraining the noise model with a reconstruction task before using it for the importance map extraction task, we facilitate and accelerate the training of our U-noise model, since the network has already learned features relating to the structure of the brain.

The reconstruction model consists of an encoder and a decoder part. It uses as input the 3D MR volumes and its task is to reconstruct the volumes as well as possible. It is trained with a voxel-wise MSE loss. The number of blocks used, meaning the number of downsample and upsample layers, is 5, while the number of output channels after the first layer is 16.

\subsubsection{Brain Age Importance Map Extraction}

For the extraction of the importance maps, we extend the U-noise architecture \cite{Koker} to 3D. The idea behind U-noise is that if one voxel is important for the prediction task, in our case brain age regression, the addition of noise on it will harm the performance of the prediction model. On the other hand, if a voxel is not relevant for the task, the addition of noise will not affect the performance.

More specifically, after the pretraining phase with the reconstruction loss, the 3D U-Net, i.e. noise model, is used for the task of extraction of a 3D noise mask that provides a noise level for every voxel of the input 3D image. A sigmoid function is applied so the values of the mask are between 0 and 1. The values are scaled to [$v_{min}$, $v_{max}$], where $v_{min}$, $v_{max}$ are hyperparameters. The rescaled mask is then multiplied by $\epsilon \sim \mathcal{N}(0,1) $, which is sampled from the standard normal distribution and the output is added to the input image element-wise in order to extract the noisy image. Given an image $X$, its noisy version can be given by Equation \eqref{eq:1}: 
\begin{equation}
X_{noisy} = X + f_{\psi}(X)(v_{max}-v_{min})\epsilon + v_{min}, 
\label{eq:1}
\end{equation}
We then use the noisy image as input for the already trained prediction model with frozen weights, and see how it affects its performance. The purpose is to maximise the noise level in our mask, while simultaneously keeping the performance of our prediction model as high as possible. In order to achieve that, we use a loss function with two terms, the noise term, which is given by $-log(f_{\psi}(X))$ and motivates the addition of noise for every voxel, and the prediction term, which is an MSE loss whose purpose is to keep the prediction model unchanged. The two loss terms are combined with a weighted sum, which is regulated by the ratio hyperparameter $r$. It is important to note that at this stage the parameters, $\theta$, of the prediction model, $f_{\theta}$, are frozen and it is not being trained. Instead, the loss function $\mathcal{L}$, is driving the training of the noise model $f_{\psi}$. Given an input image $X$ and label $y$, the loss function $\mathcal{L}$ takes the form shown in Equation \eqref{eq:2}

\begin{equation}
\mathcal{L} = \underbrace{(f_{\theta}(X) - y)^2}_\text{prediction term} - r \underbrace{log(f_{\psi}(X))}_\text{noise term}
\label{eq:2}
\end{equation}
The values $v_{min}$, $v_{max}$, as well as $r$ are hyperparameters and are decided based on the performance of our noise model on the validation set.

\section{Results}

\begin{figure}[t]
\centering
\includegraphics[width = \textwidth]{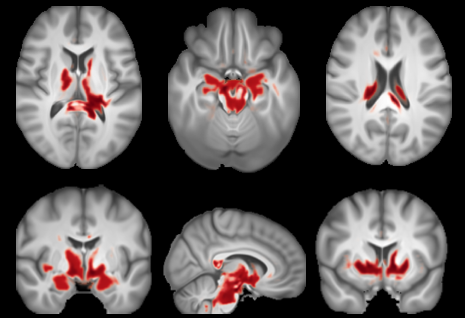}
\caption{Six different slices of the population-based importance map on top of the normalised average of the brain MR Images of the test set. The most important parts of the brain for the model's prediction are highlighted in red. The most significant regions for the task of brain aging are mesial temporal structures including the hippocampus, brainstem, periventricular and central areas. The results are in agreement with the relevant literature and previous studies.} \label{fig2}
\end{figure}

From the 13,750 3D brain images, 75\% are used for the training set, 10\% for the validation set and 15\% for the test set. All the networks are trained with backpropagation \cite{Rumelhart1986LearningErrors} and adaptive moment estimation (Adam) optimizer \cite{Kingma2015Adam:Optimization} with initial learning rate lr=0.0001, reduced by a factor of 10 every 10 epochs. The experiments are implemented on an NVIDIA Titan RTX using the Pytorch deep-learning library \cite{Paszke2019PyTorch:Library}. 

\subsection{Age Estimation}
We train the prediction model for 40 epochs using a batch size of 8 using backpropagation. We use MSE between the chronological and biological age of the subject as a loss function. The model achieves a mean absolute error (MAE) of about 2.4 years on the test set.

\subsection{Population-based importance maps}
The noise model is trained for 50 epochs and with a batch size of 2, on four GPUs in parallel. Different values were tested for hyperparameters $v_{min}$, $v_{max}$ and $r$. The chosen values, for which the network performed the best in the validation set, are $v_{min}=1$, $v_{max}=5$ and $r=0.1$. We average the importance maps for all the subjects of the test set, ending up with a population-based importance map. We use a threshold in order to keep only the top 10\% of the voxels with the lowest tolerated noise levels, meaning the most important ones for brain aging. We then apply a gaussian smoothing filter with a kernel value of 1. Different slices of the 3D population-based importance map are shown in Figure \ref{fig2}. As it can be seen, the areas that are the most relevant for brain aging according to the model's predictions are mesial temporal structures including the hippocampus, brainstem, periventricular and central areas.  

\section{Discussion}

Understanding the logic behind a model’s decision is very important in assessing trust and therefore in the adoption of the model by the users \cite{Gilpin2019}. For instance, the users should be assured that correct predictions are an outcome of proper problem representation and not of the mistaken use of artifacts. For this reason, some sort of interpretation may be essential in some cases. In the medical domain \cite{Holzinger2017}, the ability of a model to explain the reasoning behind its predictions is an even more important property of a model, as crucial decisions about patient management may need to be made based on its predictions.

In this work, we explored which parts of the brain are important for aging. In order to do so, we made the assumption that unimportant voxels/parts of the brain are not useful for brain age estimation and are not utilized by the prediction model. We trained a prediction model, which accurately estimated brain aging, and a noise model, whose purpose is to increase the noise in the input images voxel-wise, while also keeping the performance of the prediction model unaffected. As can be seen from Figure \ref{fig2}, our importance maps are in agreement with the existing neuropathology literature \cite{Savva2009}. More specifically, it is shown that the hippocampus and parts of the ventricles are where the prediction model focuses to make its decisions. 

On the other hand, the differences in the cerebral cortex appear to not be getting captured by the network. In our understanding, there are two reasons behind this. Firstly, the age range of the subjects (44-73 years old) is not large enough for the network to make conclusions. At the same time, the changes in the cerebral cortex are more noticeable after the age of 65 years. In our case, we probably do not have enough subjects in that age range in order to facilitate the network into capturing these differences.

The images that have been used in this study are non-linearly registered, since UKBB provides them ready for use and in the literature more works use the provided preprocessed dataset and therefore, comparison is easier to be done. However, it has been noticed that using non-linearly registered images may lead to the network's missing of subtle changes away from the ventricles, such as cortical changes \cite{Dinsdale2021}. 

In the future, a similar experiment will be conducted with linearly registered images instead of non-linearly registered ones because we believe that, although the performance of the prediction model might be slightly lower, the importance maps will not only be focused on the ventricles and the hippocampus, but also on more subtle changes in the cerebral cortex. Additionally, UKBB provides a variety of other non-imaging features, including biomedical and lifestyle measures, and we intend to test our method on related regression and classification tasks, such as sex classification. In the case of classification tasks we will be also comparing against gradient-based interpretability approaches, such as Grad-CAM \cite{Selvaraju2020} and guided backpropagation \cite{Springenberg2015}, since the setting allows for their use.

\section{Conclusion}
In this work, we extend the use of U-noise \cite{Koker} for 3D inputs and brain age regression. We use 3D brain MR images to train a prediction model for brain age and we investigate the parts of the brain that play the most important role for this prediction. In order to do so, we implement a noise model, which aims to add as much noise as possible in the input image, without affecting the performance of the prediction model. We then localise the most important regions for the task, by finding the voxels that are the least tolerant to the addition of noise, which for the task of brain age estimation are mesial temporal structures including the hippocampus and periventricular areas. Moving forward, we plan to test our interpretability method on classification tasks, such as sex classification as well, and compare with gradient-based methods, which are valid for such tasks.

\subsubsection{Acknowledgements} KMB would like to acknowledge funding from the EPSRC Centre for Doctoral Training in Medical Imaging (EP/L015226/1).

\bibliographystyle{splncs04}
\bibliography{paper15}

\end{document}